\makeatletter
    \def\@copyrightspace{\relax}
\makeatother

\documentclass[sigconf]{acmart}
\AtBeginDocument{%
  \providecommand\BibTeX{{%
    \normalfont B\kern-0.5em{\scshape i\kern-0.25em b}\kern-0.8em\TeX}}}

\setcopyright{acmcopyright}
\copyrightyear{2023}
\acmYear{2023}
\acmDOI{XXXXXXX.XXXXXXX}

\acmConference[CHI '23]{The ACM CHI Conference on Human Factors in Computing Systems}{April 23-28,
  2023}{Hamburg, Germany}
\begin{document}
\settopmatter{printacmref=false}
\setcopyright{none}
\renewcommand\footnotetextcopyrightpermission[1]{}
\pagestyle{plain}

\title{Repurposing Text-Generating AI into a Thought-Provoking Writing Tutor}

\author{Tyler Taewook Kim}
\authornote{Both authors contributed equally.}
\email{tk2891@columbia.edu}
\affiliation{%
  \institution{Columbia University}
  \streetaddress{116th and Broadway}
  \city{New York}
  \state{NY}
  \country{USA}
  \postcode{10027}
}

\author{Quan Tan}
\email{qt2142@columbia.edu}
\authornotemark[1]
\affiliation{%
  \institution{Columbia University}
  \streetaddress{116th and Broadway}
  \city{New York}
  \state{New York}
  \country{USA}
  \postcode{10027}
}

\renewcommand{\shortauthors}{Kim and Tan.}

\begin{abstract}
  Text-generating AI technology has the potential to revolutionize writing education. 
  However, current AI writing-support tools are limited to providing linear feedback to users. 
  In this work, we demonstrate how text-generating AI can be repurposed into a thought-provoking writing tutor with the addition of recursive feedback mechanisms.
  Concretely, we developed a prototype AI writing-support tool called \texttt{Scraft} that asks Socratic questions to users and encourages critical thinking.
  To explore how \texttt{Scraft} can aid with writing education, we conducted a preliminary study with 15 students in a university writing class.
  Participants expressed that \texttt{Scraft}'s recursive feedback is helpful for improving their writing skills.
  However, participants also noted that \texttt{Scraft}'s feedback is sometimes factually incorrect and lacks context. 
  We discuss the implications of our findings and future research directions.
\end{abstract}





\settopmatter{printacmref=false}
\maketitle

\section{Introduction} 
Before the internet, conducting information search for writing was an analog and resource-intensive process. 
However, with the widespread adoption of the internet, this process underwent a great change, giving rise to the information age and the democratization of knowledge.
We believe that text-generating AI will lead to an even more radical transform to writing, by fundamentally transforming both the writing process and writing education.

One's writing skills may soon depend on one's ability to use AI writing-support tools, analogous to how information search skills depend on one's ability to use the internet.
While this may exhibit efficacy in writing productivity, this presents implications in writing education, as AI may soon contribute to a shift in core writing pedagogies such as critical thinking and self-reflection. As such, we contend that AI writing-support tools should focus on preserving writing's recursive and thought-provoking nature.

\section{The importance of recursive feedback in writing education} 

Current AI writing-support tools are typically limited to providing linear feedback to users. They facilitate a writing process where text-generating AI directly provide outputs to user prompts, greatly minimizing opportunities for recursive and iterative user involvement.
We are particularly concerned that excessive exposure to such tools may deprive users of the opportunity to develop their writing skills, including skills that are often co-developed with writing such as critical thinking.

This is because writing is an inherently recursive process, rather than a linear one. 
Thus, we contend it is critical that educational AI writing-support tools preserve the recursive nature of writing. 
For instance, the process of argument development and feedback reception is a key element of writing. 
To preserve this process, AI writing-support tools can utilize recursive feedback, e.g., continuous generation of Socratic questions, as opposed to linear feedback, e.g., direct answer generation.
In this work, we focus on providing recursive feedback in the form of Socratic questions, however, recursive feedback refers to the general cyclical process in which a user’s understanding serves as input for an iterative exchange that continues to refine the learning experience~\cite{Okita2013}.

AI writing-support tools that are limited to linear feedback only warrant 'regenerations' or edits.
However, tools with recursive feedback mechanisms can enable users to improve their thinking and writing organically as the feedback is tailored to users' individual writing experience.
Throughout the writing process, users can request continuous and personalized feedback from the AI writing-support tool, i.e., AI can be a personal writing tutor.

\begin{figure*}[t!]
\centering
\includegraphics[width=0.95\linewidth]{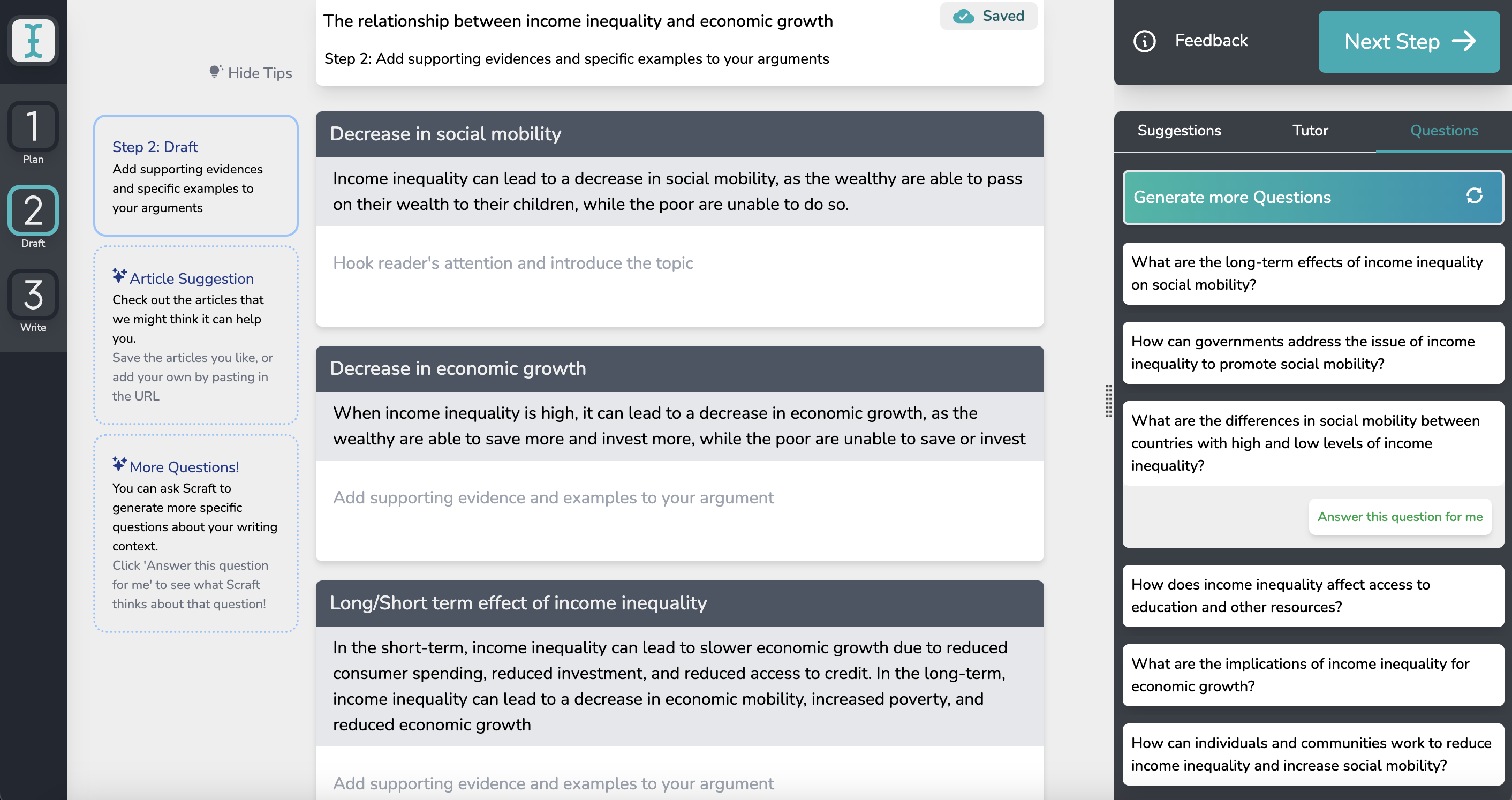} 
\caption{\texttt{Scraft} is an AI writing-support tool that provides recursive feedback to users in the form of thought-provoking Socratic questions. More information is available at \url{https://scraft.ai}.}
\label{fig:scraft}
\end{figure*}

 \section{\texttt{Scraft}: Thought-provoking AI writing tutor}

To realize its idea, we developed a prototype AI writing-support tool called \texttt{Scraft} (Figure \ref{fig:scraft}) that provides recursive feedback based on the Socratic questioning method.
The impetus for exploring this approach lies in its capacity to cultivate critical thinking, which serves as a fundamental element of writing pedagogy. Socratic questioning fosters a disciplined mode of thought that actively monitors, evaluates, and restructures cognitive processes, emotions, and actions in a rational manner~\cite{Elder1998}.

The current prototype is designed to generate elementary recursive feedback by continuously producing Socratic questions as users engage in the writing process.
When users incorporate the provided Socratic questions into their work, \texttt{Scraft} adapts to this new input and generates an updated set of feedback. 
This functionality is integrated within a conventional writing document interface, featuring a sidebar for displaying the generated questions. Additionally, users have the option to utilize an "answer the question for me" button, which enables them to receive suggested responses to the Socratic questions at any given point during the writing process.

\section{Preliminary study}
To explore how \texttt{Scraft} can aid with writing education, we conducted a preliminary study with 15 students in a university writing class.
Participants used \texttt{Scraft} for a two-week period to work on a class assignment.
Afterwards, we conducted 15 minutes interviews with eight students.
Below we summarize the participants' praises and complaints on the current prototype.

Starting with praises, many participants said they experienced an improvement in their writing skills, and attributed it to \texttt{Scraft}'s Socratic questions. 
These participants said that the questions encouraged them to critically think about aspects of their writing they had not previously considered.
\texttt{Scraft}'s recursive feedback pinpointed areas in the participants' writing that require further development.
By doing so, it encouraged the participants to conduct more research on the writing topic and develop a deeper understanding.
Participants also mentioned that \texttt{Scraft}'s Socratic questions were useful for anticipating potential counterarguments to their claims and preemptively strengthen their writing.

The main complaint on the current prototype was that the provided Socratic questions varied in their usefulness.
Participants said some questions lacked relevance to the writing topic.
In other instances, the provided feedback contracted one another or lacked factual accuracy.
We find this to be consistent with the findings of Dikli and Blyele~\cite{DIKLI20141,Dikli_2013}: Compared to human feedback, automated essay feedback may often yield mixed outcomes as the effectiveness of such feedback may be undermined by factors such as incomprehensibility, overly general content that is challenging to implement, and overall inaccuracies in identifying essay sections with errors or requiring correction.

\section{Conclusion}
In summary, we demonstrated how text-generating AI can be repurposed into a thought-provoking writing tutor by providing recursive feedback to users in the form of Socratic questions.
Our preliminary study with \texttt{Scraft} show the potential of this approach.
However, it also revealed limitations: Socratic questions generated by current text-generating AI technology sometimes lack relevance to the writing topic and are factually incorrect.
These findings underscore the need for further research on educational AI writing-support tools to better support the recursive nature of the writing process and uphold pedagogical principles that foster critical thinking and self-reflection in student writers.
We argue our proposed approach not only maintains but enhances the natural processes of writing, i.e. drafting, which in culmination, fosters the core pedagogical aspects of writing education.

\begin{acks}
We thank our participants for generously sharing their time and experiences. We also thank Katy Ilonka Gero, Sunnie S. Y. Kim, and the anonymous reviewers for their thoughtful feedback.
\end{acks}

\bibliographystyle{ACM-Reference-Format}
\bibliography{sample-base}

\end{document}